\documentclass[10pt,usenatbib]{article}

\usepackage{amsfonts}

\usepackage{amssymb}

\usepackage{amsmath}

\usepackage{longtable,array}
\usepackage{graphicx}
\usepackage{comment}

\usepackage{bm}

\begin{document}

\newcommand{\half}{\mbox{$\textstyle \frac{1}{2}$}}
\newcommand{\ket}[1]{\left | \, #1 \right \rangle}
\newcommand{\bra}[1]{\left \langle #1 \, \right |}
\newcommand{\beq}{\begin{equation}}
\newcommand{\eeq}{\end{equation}}
\newcommand{\bea}{\begin{eqnarray}}
\newcommand{\eea}{\end{eqnarray}}
\newcommand{\req}[1]{Eq.\ (\ref{#1})}
\newcommand{\gcc}{{\rm~g\,cm}^{-3}}
\newcommand{\Compton}{\lambda\hspace{-.44em}\raisebox{.6ex}{\mbox{-$\!$-}}%
\raisebox{-.3ex}{}_{\hspace{-1pt}{\mbox{$_\mathrm{C}$}}}}
\newcommand{\kB}{k_\mathrm{B}}
\newcommand{\omc}{\omega_\mathrm{c}}
\newcommand{\omg}{\omega_\mathrm{g}}
\newcommand{\mel}{m_e}
\newcommand{\xr}{x_{\rm r}}
\newcommand{\EF}{\epsilon_{\rm F}}
\newcommand{\Ne}{{\cal N}_B(\epsilon)}
\newcommand{\Necl}{{\cal N}_0(\epsilon)}
\newcommand{\dfde}{{\partial f^{(0)} \over\partial\epsilon}}
\newcommand{\am}{a_\mathrm{m}}
\newcommand{\dd}{{\rm\,d}}
\newcommand{\vB}{\bm{B}}
\newcommand{\oc}{op.\ cit.~}
\newcommand{\dotZ}{\mbox{$\dot{\mbox{Z}}$}}

\newcommand{\msun}{\mbox{$M_\odot$}}
\renewcommand{\thefootnote}{\fnsymbol{footnote}}

\begin{center}
{\Large \bf Lev Landau and the conception of neutron stars}
\end{center}

\begin{center}
Dmitry~G. Yakovlev$^1$,
Pawe{\l}~Haensel$^2$,
Gordon Baym$^3$,
Christopher~J.~Pethick$^{4,5}$ \\
{\small \it $^1$Ioffe Physical Technical Institute, Politekhnicheskaya
26, 194021
St.-Petersburg, Russia}\\
{\small \it $^2$N.~Copernicus Astronomical Center, Bartycka 18,
00-716 Warsaw, Poland}\\
{\small \it $^3$Department of Physics, University of Illinois, 1110
W. Green Street, Urbana, Illinois 61801, USA}\\
{\small \it $^4$Niels Bohr International Academy, Niels Bohr
Institute,
Blegdamsvej 17, DK-2100, Copenhagen, Denmark}\\
{\small \it $^5$ NORDITA, Royal Institute of Technology and Stockholm University,\\
Roslagstullsbacken 23, SE-10691 Stockholm, Sweden }
\end{center}



\begin{abstract}
We review the history of neutron star physics in the 1930s that is
related to L.\ Landau. According to recollections of Rosenfeld (1974,
Proc.\ 16th Solvay Conference on Physics, p.\ 174), Landau improvised
the concept of neutron stars in a discussion with Bohr and Rosenfeld
just after the news of the discovery of the neutron reached
Copenhagen in February 1932. We present arguments that the discussion
took place in March 1931, before the discovery of the neutron, and
that they in fact discussed the paper written by Landau in Zurich in
February 1931 but  not published until February 1932 ({\it Phys.\ Z.\
Sowjetunion} {\bf 1} 285). In his paper Landau mentioned the possible
existence of dense stars which look like one giant nucleus; this can
be regarded as an early theoretical prediction or anticipation of
neutron stars, prior to the discovery of the neutron. The coincidence
of the dates of the neutron's discovery and the paper's publication
has led to an erroneous association of the paper with the discovery
of the neutron. In passing, we outline the contribution of Landau to
the theory of white dwarfs and to the hypothesis of stars with
neutron cores.
\end{abstract}

\def\baselinestretch{1}

\section{Introduction}
\label{s:introduct}

\renewcommand{\thefootnote}{\arabic{footnote}}

\newcommand{\st}{(1)}

Neutron stars, highly compact objects with a mass of order that of
the Sun but only about 10 km in radius, are the engines of many
remarkable astrophysical phenomena, from radio pulsars to compact
X-ray sources. These objects are not only fascinating in themselves,
but they have been the basis of critical tools in astrophysical
measurements; in addition they provide a window into the properties
of matter at the highest densities. Although neutron stars were first
discovered as radio pulsars in 1967 \cite{Hewish}, they had been the
subject of theoretical investigations since the 1930s, and had been
studied in the hope of resolving fundamental questions of the final
states of stellar evolution, the source of the energy released in
supernovae, and even the energy source of ordinary stars. In these
initial developments the great Soviet physicist, Lev D.\ Landau
(1908--1968), played a controversial role. Our focus in this paper is
to clarify Landau's contributions to neutron stars while summarizing
the early history of the physics that fed into the idea of neutron
stars. A brief summary of this investigation was given in Ref.\
\cite{NSB}.

To set the stage, we recall an oft-quoted reminiscence of L\'eon Rosenfeld at the 1973 Solvay Conference. He described
a discussion between Niels Bohr,  Landau, and himself in Copenhagen
in early 1932 \cite{Rosenfeld}:

\begin{quote}
I recall when the news on the neutron's discovery reached Copenhagen,
we had a lively discussion on the same evening about the prospects
opened by this discovery. In the course of it Landau improvised the
conception of neutron stars -- `unheimliche Sterne', weird stars,
which would be invisible and unknown to us unless by colliding with
visible stars they would originate explosions, which might be
supernovae. Somewhat later, he published a paper with Ivanenko in
which he again mentioned neutron stars as systems `to which quantum
mechanics would not be applicable.'
\end{quote}
This story was first repeated by Baym and Pethick \cite{bp75}, and
later by Baym \cite{Baym}, as well as by Shapiro and
Teukolsky in their monograph \cite{st83} (first footnote on p.\ 242). It naturally
created the impression, widely mentioned in the literature, that
neutron stars were predicted by Landau in 1932 as a direct
consequence of the discovery of the neutron. The historical facts do
not bear the story out.

\begin{figure}[tbh]
\begin{center}
\includegraphics[width=7 cm]{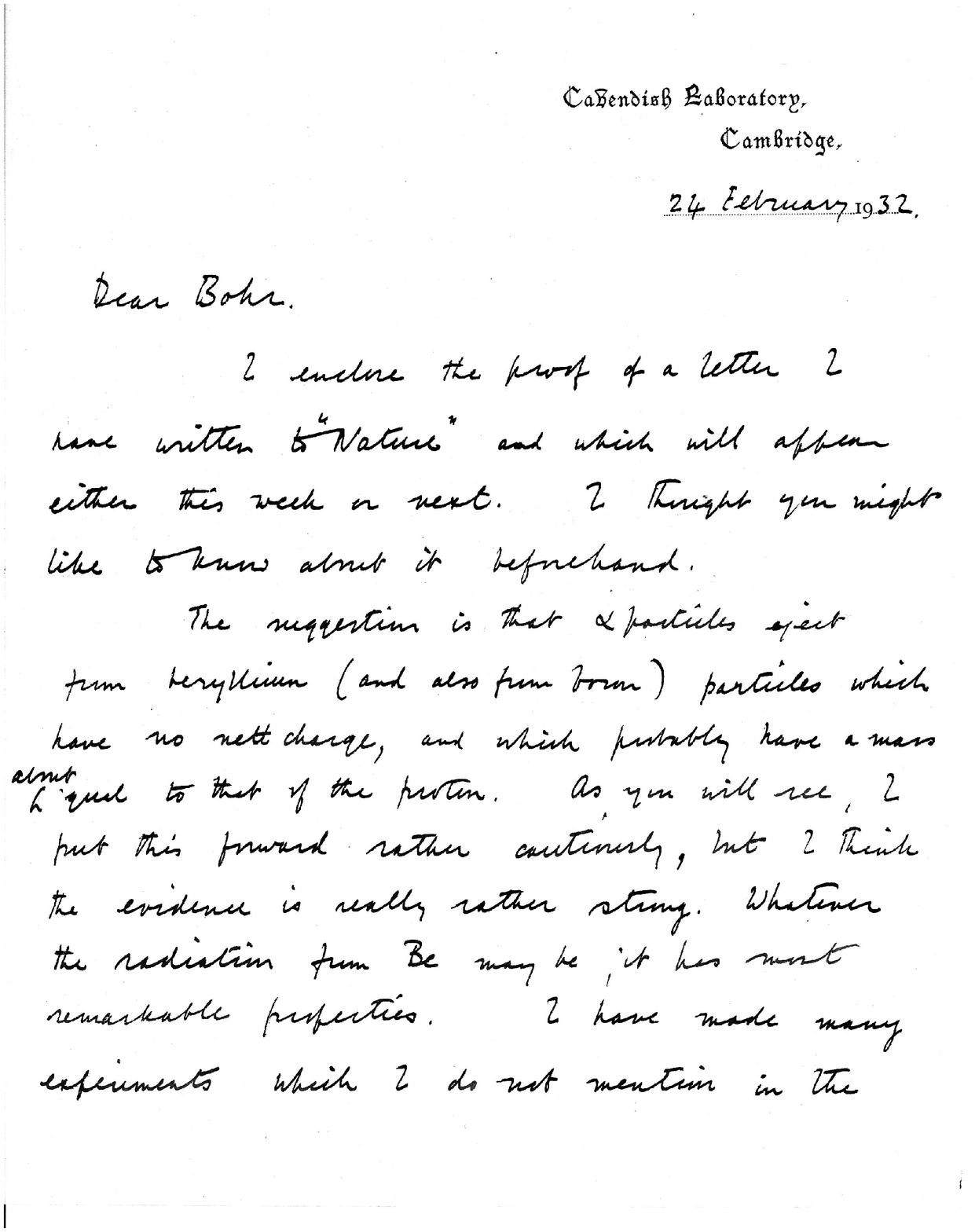}
\includegraphics[width=7 cm]{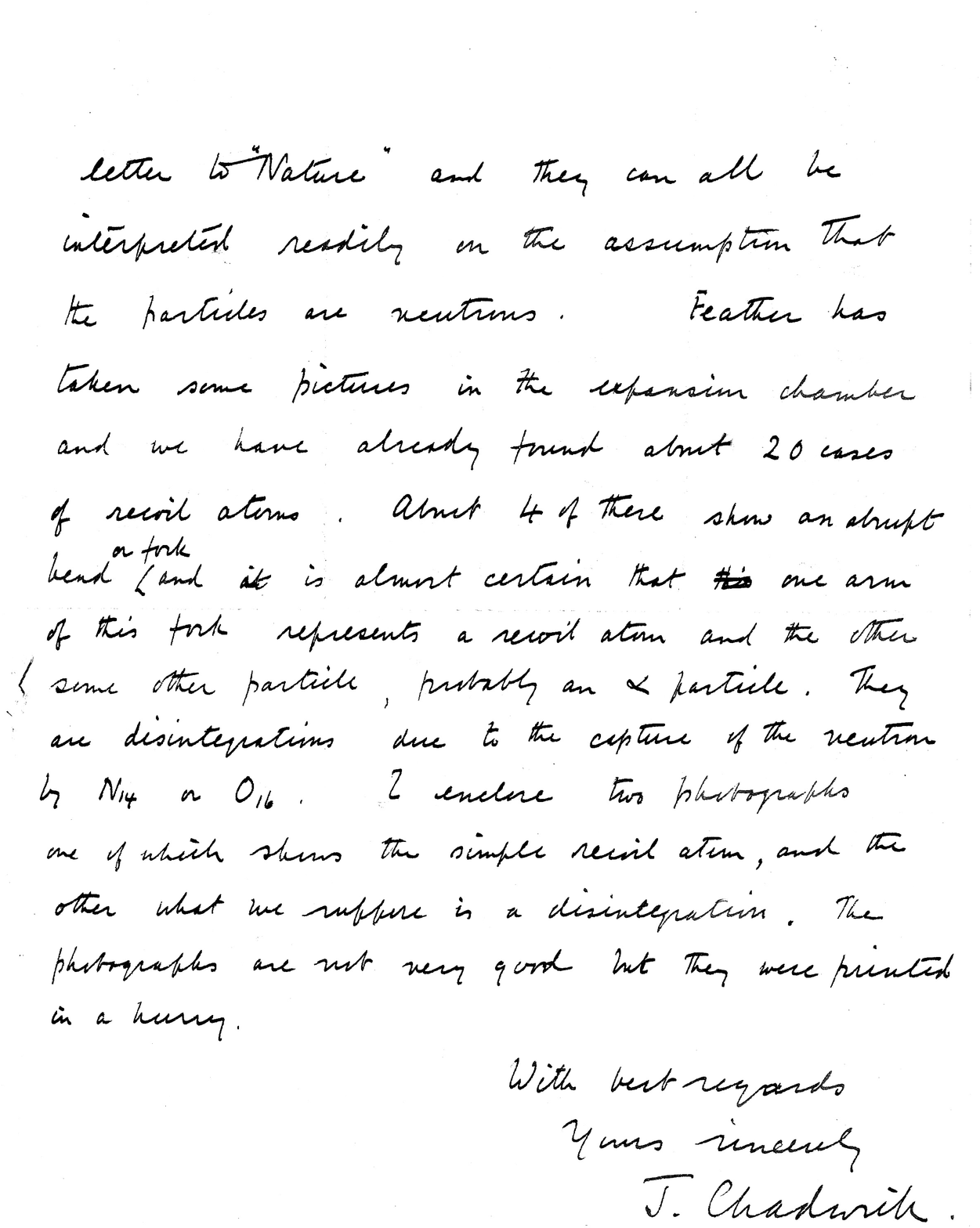}
\caption{\small Chadwick's letter to Bohr on the discovery of the
neutron. Courtesy  of the Niels Bohr Archive, Copenhagen. Reproduced
with the kind permission of Judith Chadwick.}
\label{fig:Chadwick}
\end{center}
\end{figure}

James Chadwick, who discovered the neutron in February 1932, reported
his finding to Bohr in a letter dated 24 February 1932
(Fig.~\ref{fig:Chadwick}) which would date this purported evening
discussion to shortly afterwards.  However, as is often the case in
recollections by participants many years after the fact, Rosenfeld's
account is only a fuzzy hint of the actual story. On the one hand,
Landau was not in Copenhagen in 1932.  Furthermore, neither were Bohr
nor Rosenfeld in Copenhagen when Chadwick's letter arrived; both were
travelling, and Chadwick's letter apparently lay on Bohr's desk until
Bohr returned in the second half of March.  As we discuss below in
Sec.\ \ref{Landau's paper `On the theory of stars'}, Landau had
already had the idea of the possibility of matter in a star ``forming
one gigantic nucleus,'' a year earlier. In addition he never wrote a
paper with the Soviet physicist Dmitry Ivanenko on stars (e.g., Ref.\
\cite{Aksenteva}); most likely Rosenfeld was thinking of Landau's
earlier paper (Sec.\ \ref{Revisiting Rosenfeld's recollection}).
Nonetheless, Rosenfeld's recollection does provide a useful stepping
stone into the early history of neutron stars.

\section{Landau and Bohr}
\label{Landau and Bohr}

Landau's trajectory through Europe and his interactions with Bohr are
crucial to understanding the correct history of Landau's connection
to neutron stars.  Landau  graduated from Leningrad State University
in January 1927, at the age of 19,  and was appointed to a position,
resembling that of a present day graduate student, at Leningrad
Physical Technical Institute (now the Ioffe Physical Technical
Institute, St.\ Petersburg). His supervisor was Yakov (or Jakob)~I.\
Frenkel, head of  the Theoretical Physics Department. A brilliant
student, Landau was awarded one of two stipends of the Ministry of
Education (the `Narodnyi Komis\-sariat Pros\-vya\-stche\-niya', or
the `People's Commissariat of En\-light\-en\-ment') for a  one and a
half year trip abroad for scientific work.

\begin{figure}[tbh]
\begin{center}
\includegraphics[width=5.0cm]{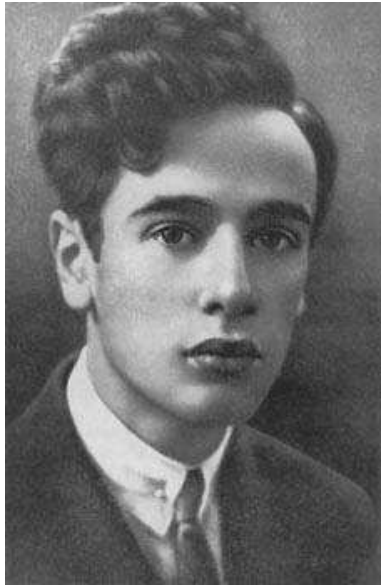}
\caption{\small Lev Landau in 1929. 
}
\label{fig:Landau}
\end{center}
\end{figure}

\begin{figure}[tbh]
\begin{center}
\includegraphics[width=10.0cm]{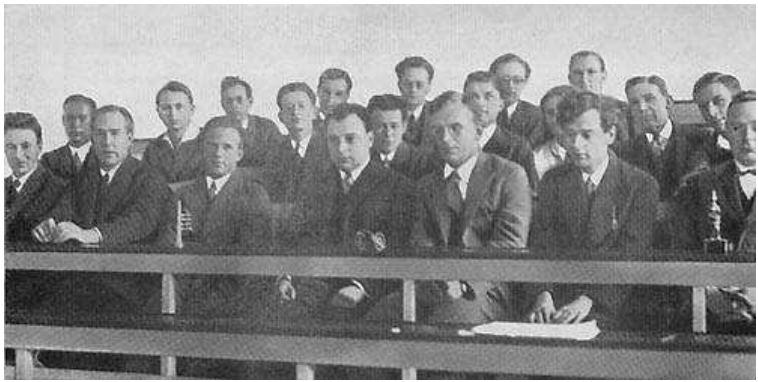}
\caption{\small The annual conference at the Institute for
Theoretical Physics in Copenhagen in 1930. First row (from left to
right): O.\ Klein, N.\ Bohr, W.\ Heisenberg, W.\ Pauli, G.\ Gamow,
L.\ Landau, H.\ Kramers. Courtesy of the Niels Bohr Archive,
Copenhagen.}
\label{fig:seminar}
\end{center}
\end{figure}

Landau, seen in Fig.~\ref{fig:Landau} at the time (from Ref.\
\cite{Bessarab}), started his trip in October 1929, visiting numerous
physics centers in Europe including (chronologically) Berlin,
G\"ottingen, Leipzig, Zurich, Copenhagen, Cambridge, and
Copen\-hagen, Zurich, and Copenhagen again. In particular, he visited
Bohr in Co\-pen\-ha\-gen 8 April -- 3 May 1930, 20 September -- 20
November 1930, and 25 February -- 19 March 1931 (Ref.\
\cite{Pais1991}, footnote on p.\ 359). Landau liked Bohr and looked
upon him as his only teacher; Bohr returned the friendship. In the
beginning of 1931, when Landau's Soviet stipend had run out, Bohr
helped him obtain a Rockefeller Fellowship which allowed Landau to
prolong his stay and visit Bohr in Copenhagen the third time. Figure
\ref{fig:seminar} is a photo of Niels Bohr's annual conference in
Copenhagen in 1930 with Landau in the front row on the right. After
the third visit Landau returned to Leningrad and moved to Kharkov the
same year.

\newcommand{\Paisa}{(25)}

Landau and Bohr met again in Copenhagen and the USSR.   Landau
visited Bohr in Copenhagen 18 September -- 3 October 1933  and 1 June
-- 8 July 1934, to attend conferences organized by Bohr. After 1934,
Landau did not travel abroad (except for medical treatment in
Czechoslovakia in 1963 after his terrible car accident on 7 January
1962, which terminated his scientific career). Bohr visited the USSR
and met with Landau there three times: in May 1934 (Leningrad,
Moscow, Kharkov), June 1937 (in Moscow, during Bohr's half-year tour
to Japan, the USA, China, and the USSR), and May 1961 (Moscow) -- see
Ref.\ \cite{Pais1991},  pp.\ 415,  417, and 528.

\section{Landau's paper `On the theory of stars'}
\label{Landau's paper `On the theory of stars'}

On 7 January 1932,  more than one month before Chadwick's
announcement of the discovery of the neutron, and even before
Chadwick started his successful experiment (see Sec.\ \ref{The
discovery of the neutron}), Landau submitted a paper \cite{Landau32}
in English to the  {\it Physikalische Zeitschrift der Sowjetunion},
the first Soviet physical journal published in languages other than
Russian. The paper was published on 29 February 1932, less than a
week after Chadwick's letter to Bohr. Interestingly, the last line of
the published paper says: `February 1931, Zurich,' from which one
would conclude that Landau wrote the paper one year before the
discovery of the neutron, in Zurich, but for unknown reasons had
delayed sending it to a journal for nearly a year.\footnote{Karl
Hufbauer \cite{Hufbauer} speculates that, ``He probably intended to
publish it in {\em Nature}, the field's leading popular venue. As
things turned out, however, he ended up carrying the manuscript back
to the Leningrad Physico-Technical Institute when he returned to the
Soviet Union that spring and submitted it to the new {\em
Physikalische Zeitschrift der Sowjetunion} early the next year.''}
The omission of the February 1931 date in reprints of Landau's paper
\cite{rep1,rep2}, aided by Rosenfeld's recollection, has led to an
erroneous association, of Chadwick's discovery with Landau's early
ideas on dense stars (see Sec.\ \ref{Revisiting Rosenfeld's
recollection}).

The four page paper, written in Landau's concise style, consists of
two parts, both of which are significant.  The first part is devoted
to an incisive calculation of the maximum mass of a white dwarf, and
the second to speculations on the structure and physics of stars of
higher density. The maximum mass  is the so-called Chandrasekhar mass
limit of white dwarfs, the highest mass supportable against gravity
by the pressure of relativistic degenerate electrons. The history of
the theoretical prediction of this mass is described, for instance,
by Nauenberg \cite{Nauenberg} and Shaviv \cite{Shaviv}. The problem
was a forefront application of quantum statistical mechanics to the
open question of the nature of white dwarfs, but it is not clear to
what extent Landau was aware of the developments that were taking
place.\footnote{The first estimates of a maximum mass were made by
Wilhelm Anderson \cite{Anderson} and then by Edmund Clifton Stoner
\cite{Stoner}. The exact calculation of the maximum mass was first
carried out by Chandrasekhar in July 1930, without knowledge of
Anderson's and Stoner's works, and was published in Ref.\
\cite{Chandra1}.} Landau calculated the maximum mass independently
but later than Chandrasekhar.  He presents an important simple
formula, \
\begin{equation}
   M_0=\frac{3.1}{m^2} \left(\hbar
   c \over G\right)^{3/2},
\end{equation}
relating the maximum mass to four physical parameters -- Planck's
constant $\hbar$, the speed of light $c$, Newton's gravitational
constant $G$, and $m$, the mass of the matter per electron. For white
dwarf interiors one has $m \approx 2 m_p$, with $m_p$ the proton
mass, which gives the well known value $M_0 \approx 1.44\,M_\odot$,
where $M_\odot$ is the mass of the Sun. Note that Landau gives
$M_0\approx 1.5\,M_\odot$ (while Chandrasekhar \cite{Chandra1}
originally obtained $M_0\approx 0.91\,M_\odot$ because he assumed an
unreasonably high value of $m=2.5\,m_p$, as explained, e.g., in
Ref.~\cite{Shaviv}). Landau did not state it explicitly, but his
formula shows that the scale of the mass of white dwarfs is given by
the proton mass, $m_p$, times $\alpha_G^{-3/2}\approx 10^{57}$, where
$\alpha_G = m_p^2 G/\hbar c$ is the modern `gravitational fine
structure constant'.

Landau next discusses more massive stars. Although he would not write
on neutrons in stars until nearly six years later, he was confronting
at this time the theoretical question of the collapse of stars of
mass greater than $M_0$ to higher and higher densities, and was
toying with ideas of heavier stars having cores in which even ``the
laws of ordinary quantum mechanics break down\ldots''.

Landau's colleague and friend, Matvei P.\ Bronstein, a remarkably
talented theoretical physicist with very broad interests from
solid-state physics to quantum gravity (and sadly shot in 1938 by the
Stalin regime), expanded \cite{Bronstein} on how Landau was thinking.
While the problem of the source of energy in stars was unsolved at
the time, the great difference between ordinary main sequence stars
and white dwarfs was not clear. To Landau the energy problem appeared
to be an obstacle that demanded new physical theories going beyond
standard quantum mechanics. Landau (wrongly) assumed that the maximum
mass limit, $1.5\,M_\odot$, he calculated was the maximum for all
stars to which standard physics was applicable. The possibility that
massive ordinary stars were composed of low-density non-degenerate
matter was not taken into account. Furthermore, observational
evidence on the existence of many massive stars was limited. Any star
with $M$ greater than $1.5\,M_\odot$ was treated as `pathological,'
containing a core of very dense matter, where standard physical laws
(including energy conservation) were violated; he also assumed that
dense `pathological' regions could appear in less massive stars.
Landau writes,

\begin{quote}
As we have no reason to believe that stars can be divided into two
physically different classes according to the condition $M>$ or $<
M_0$, we may with great probability suppose that all stars possess
such pathological regions. It does not contradict the above
arguments, which prove only that the condition $M > M_0$ is
sufficient (but not necessary) for the existence of such regions. It
is very natural to think that just the presence of these regions
makes stars stars. But if it is so, we have no need to suppose that
the radiation of stars is due to some mysterious process of mutual
annihilation of protons and electrons, which was never observed and
has no special reason to occur in stars. Indeed we have always
protons and electrons in atomic nuclei very close together, and they
do not annihilate themselves; and it would be very strange if the
high temperature did help, only because it does something in
chemistry (chain reactions!). Following a beautiful idea of Professor
Niels Bohr's we are able to believe that the stellar radiation is due
simply to a violation of the law of energy, which law, as Bohr has
first pointed out, is no longer valid in the relativistic quantum
theory, when the laws of ordinary quantum mechanics break down (as it
is experimentally proved by continuous-rays spectra and also made
probable by theoretical considerations \cite{LP31}). We expect that
this must occur when the density of matter becomes so great that
atomic nuclei come in close contact, forming one gigantic nucleus.
\end{quote}

Landau's understanding that protons and electrons constitute atomic
nuclei and do not annihilate there provides additional proof that the
paper was conceived before the discovery of the neutron. While the
suggestion that sources of stellar energy were located in
`pathological cores' was convenient, it was naive. Furthermore, that
the energy output of stars could be attributed to a breakdown of
energy conservation in quantum mechanics, while a plausible
conjecture at the time, is, as we now know, wrong.  But Landau's
anticipation of dense (neutron) stars which look like giant atomic
nuclei was prescient.

\section{The discovery of the neutron}
\label{The discovery of the neutron}

The neutron, which was predicted as early as 1920 by Rutherford
\cite{Rutherford}, was finally discovered by Chadwick
\cite{Chadwick}, who left a colorful description of his discovery
\cite{Chadwick1}. After Rutherford's prediction Chadwick had
attempted to find evidence of the neutron many times but invariably
failed. At the beginning of 1932 he became inspired by the paper of
Curie and Joliot \cite{CurieJoliot}, published on 28 January. As
acknowledged in Chadwick's second publication on the neutron's
discovery \cite{Chadwickb}, experiments similar to that of Curie and
Joliot were conducted also by Bothe and Becker \cite{Bothe}, Curie
\cite{Curie} and Webster \cite{Webster}, all of whom reported results
of bombardment of beryllium nuclei by alpha particles produced by a
polonium source. They found that neutral reaction products (which
they interpreted as gamma rays) knocked out protons from paraffin.
Chadwick and Rutherford did not believe this interpretation. Chadwick
immediately went to work, conducting a similar experiment with better
equipment and in a few days of strenuous work found the same result
\cite{Chadwick1}. However, he interpreted it correctly, as production
of neutrons (not gamma-rays) in $\alpha$+Be collisions. Finally
successful, he could even estimate, using energy and momentum
conservation in the knock-out reaction, the mass of the neutron,
which he found to be close to the proton mass. His discovery
\cite{Chadwick}, announced in {\it Nature}, was received by the
editor on 17 February  and published on 27 February 1932. More
detailed explanations were given in the next paper \cite{Chadwickb}.

Note that the first theoretical model of atomic nuclei as composed of
protons and neutrons was suggested in April 1932 by Ivanenko
\cite{Dimus}.

\section{The prediction of neutron stars after the neutron's discovery}
\label{s: Baade and Zwicky}

The first explicit prediction of neutron stars was made by Walter
Baade and Fritz Zwicky at Caltech in December 1933, nearly two years
after the discovery of the neutron, in trying to explain the enormous
energy released in supernova explosions. Reporting their results at
the meeting of the American Physical Society at Stanford on 15--16
December 1933, they wrote with remarkable foresight \cite{bz34a},
``with all reserve we advance the view that supernovae represent the
transition from ordinary stars into neutron stars, which in their
final stages consist of closely packed neutrons.'' They amplified
this indeed accurate picture in the second and third of their papers
\cite{bz34a1,bz34a2} of the following March which proposed cosmic-ray
production in supernovae, ``Such a star may possess a very small
radius and an extremely high density. As neutrons can be packed much
more closely than ordinary nuclei and electrons, the 'gravitational
packing' energy in a cold neutron star may become very large, and,
under certain circumstances far exceed the ordinary nuclear packing
fractions. A neutron star would therefore represent the most stable
configuration of matter as such.'' Only through such close packing
could they understand the energy release in supernovae which they
estimated to be equivalent to the annihilation of the order of
several tenths of a solar mass. However, they could only guess at the
scenario for forming neutron stars.  In their fourth publication
Baade and Zwicky \cite{bz34b} naively (and wrongly) explain that,
``\ldots neutrons are produced on the surface of an ordinary star''
[under the effect of cosmic rays] and ``\,`rain' down towards the
center as  we assume that the light pressure on neutrons is
practically zero.''

\begin{figure}[t]
\begin{center}
\includegraphics[width=8.0cm,angle=0,bb= 60 55 560
660]{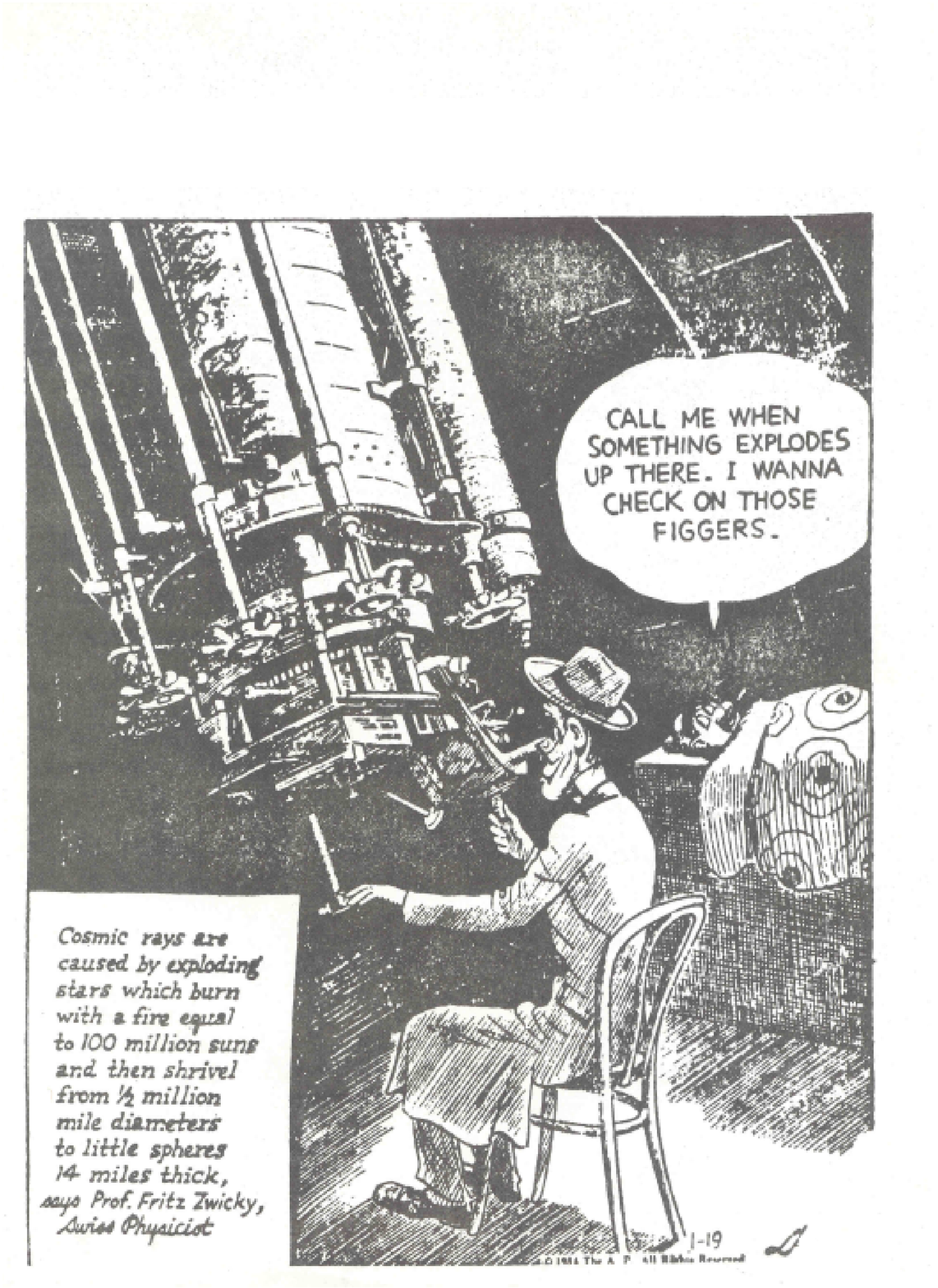} 
\caption{\small Baade and Zwicky's prediction of neutron stars,
reported in this cartoon in the \emph{Los Angeles Times} of 19
January 1934. The lower box reads: ``Cosmic rays are caused by
exploding stars which burn with a fire equal to 100 million suns and
then shrivel from 1/2 million mile diameters to little spheres 14
miles thick, says Prof.\ Fritz Zwicky, Swiss Physicist.'' Reproduced
with permission of The Associated Press, Copyright \copyright 1934.
All rights reserved.}
\label{fig:Zwicky}
\end{center}
\end{figure}

Their idea of neutron stars quickly made it into the popular press,
as Fig. \ref{fig:Zwicky} from {\it The Los Angeles Times}
illustrates.\footnote{We thank Janusz Zi{\'o}{\l}kowski for
calling this cartoon, presented in a talk of Robert P. Kirshner at
the Eight Texas Symposium on Relativistic Astrophysics
\cite{Kirshner}, to our attention. We are also grateful to R. P.
Kirshner for helpful information referring to this cartoon.}  The stellar diameter mentioned in the cartoon is
uncannily accurate.

\section{Revisiting Rosenfeld's reminiscence}
\label{Revisiting Rosenfeld's recollection}

With this history in place we ask how we can make sense of
Rosenfeld's recollection.  First, as we have seen, there was no
connection between the discovery of the neutron and Landau's concept
of dense stars.   Thus Landau clearly did not improvise the concept
of  {\it unheimliche Sterne} in a discussion with Bohr and Rosenfeld.
Most likely the discussion between Landau, Bohr, and Rosenfeld took
place during Landau's third trip to Copenhagen in February and March
1931, just after he wrote his paper in Zurich.  It was natural that
Landau at this point would discuss his paper with Bohr and Rosenfeld
(the latter had arrived in Copenhagen on 28 February 1931, see Ref.\
\cite{Rosenfeld0}), since Landau had tried to utilize Bohr's ideas of
violation of traditional physics. Then in using the term {\it
unheimliche Sterne} Rosenfeld was referring to Landau's idea of a
star as a single dense nucleus in which quantum mechanics is
violated. Once the neutron was discovered there was no longer a need
to violate quantum mechanics to avoid unobserved electron-proton
annihilations.  Given the compelling arguments that the discussion
between the three took place just after Landau's arrival in
Copenhagen in 1931, it is unlikely that it happened at a later
occasion when all three were together -- e.g., 18 September -- 3
October 1933 or 1 June -- 8 July 1934 in Copenhagen, or during one of
the two trips by Bohr to the Soviet Union in 1934 and 1937. In 1931
{\it unheimliche Sterne} would have referred simply to a star as a
single dense nucleus, more like neutron stars as we know them.

There is no evidence that Landau would have discussed {\it
unheimliche Sterne} as being ``invisible and unknown to us unless by
colliding with visible stars they would originate explosions, which
might be supernovae.''  And Rosenfeld's reference to Landau
publishing a paper with Ivanenko must have been to Landau's 1931
paper. The name of Ivanenko could well have been mentioned in
association with Ivanenko's paper \cite{Dimus} on the proton-neutron
structure of atomic nuclei written after the neutron's discovery.

\section{Stars with neutron cores}
\label{Stars with neutron cores}

The source of stellar energy remained unknown until 1938 (see, e.g.,
Ref.\ \cite{Shaviv}). In addition the ultimate fate of stars more
massive than $M_0$ remained a theoretical mystery. One suggestion was
that they developed cores of highly compressed matter (see e.g.,
Ref.\ \cite{Milne1932}). While Chandrasekhar in late 1934 at Trinity
College had given serious arguments against this possibility,
suggesting mass loss or supernova phenomena instead
\cite{Chandra1935,Chandra1935a}, Gamow \cite{Gamow37}, reviewing
Chandrasekhar's (1931) and Landau's (1932) earlier arguments, wrote
that more massive stars ``are subject to the formation of matter in
the nuclear state in their interior at some period of their
existence.'' This state would be formed by beta capture of electrons
by nuclei, changing protons to neutrons. [Although Gamow did not
refer to it, the first microscopic descriptions of the equation of
state of nuclear matter in beta equilibrium had actually been given
by Hund \cite{Hund1936} a year earlier.] Such a state would not only
provide, through gravitational release, energy ``enough to secure the
life of the star for a very long period of time,'' but eruptive
processes at the surface of the dense nuclear core could give rise to
spewing out of ``nuclear substance'' and formation of various nuclei,
thus helping with problems of nucleosynthesis.

Quite independently, Landau, addressed the origin of stellar energy
in a paper  first published in Russian in the {\it Proceedings of
USSR Academy of Sciences} \cite{Landau37}
in 1937, and then in {\it Nature} \cite{Landau38} (submitted from
Moscow via Bohr in Copenhagen in late 1937). Landau proposed that in
ordinary stars of masses greater than a critical mass $\sim
0.001\,M_\odot$, a neutron core, ``where all the nuclei and electrons
have combined to form neutrons,'' would be energetically favored over
a normal core. For Landau, it was a step from ``pathological'' cores
in his paper written in 1931 with unknown physics to neutron cores
described by standard physics. The stellar envelope contracts
(accretes) slowly onto the compact core. The huge release of
gravitational energy (due to the core compactness) in the course of
accretion would be sufficient to heat the star and support its
luminosity at the high level for a long time. In particular, Landau
finished his new paper with an appeal to build up the models of such
stellar objects. Hund's earlier work is referred to here, but only as
an afterthought at the suggestion of Bohr and M{\o}ller
\cite{Bohr1937}. Landau stressed, as did Gamow, that a neutron core
would ``give an immediate answer to the question of the sources of
stellar energy.''

We note in passing that the political situation in the Soviet Union
had turned from bad to worse in 1937, and Landau was in real danger
of arrest. To avoid this, P.~L.\ Kapitsa organized a campaign to
popularize Landau's work and make Landau a famous scientist
(including very favorable press coverage; see Hufbauer
\cite{Hufbauer} for an amusing story about the publication of the new
paper). However the political regime disregarded such campaigns.
Landau was arrested in 1938 in Moscow for `anti-Soviet activity' and
spent one year in prison. He was released (and the conditions of his
sentence were softened) mainly due to the efforts of Kapitsa who did
not fear to write several letters in Landau's defense  to Stalin and
top Soviet officials \cite{Feinberg}. Curiously, in his first letter
to Stalin on 28 April 1938, the very day of the Landau's arrest,
Kapitsa argued that Landau ``\ldots published one remarkable paper,
where he was the first to show a new source of stellar radiation.''
Thus Landau's paper was used as an argument to save his life.

When Landau was released, it was already evident (e.g., Ref.\
\cite{Bethe39}) that normal stars support their life by nuclear
burning. The problem of stellar energy was solved and Landau's appeal
was temporarily forgotten. However the idea was realized later in the
form of {\em Thorne--{\dotZ}ytkow objects}, hypothetical stellar
objects that from the outside look like ordinary giants or
supergiants. Inside, they contain large rarefied convective envelopes
and dense degenerate cores, again like giants and supergiants.
However, ordinary giants and supergiants possess degenerate electron
cores (i.e., contain white dwarfs in their centers) while these new
objects possess much denser, neutron degenerate cores (i.e., neutron
stars). The idea of constructing models of red giants or supergiants,
replacing the electron degenerate cores by compact neutron cores, was
discussed by several authors, including Fermi, Paczy\'{n}ski,
Ostriker, Bisnovatyi-Kogan, Sunyaev, and Thorne. For references, see
the paper by  Thorne and {\dotZ}ytkow \cite{ThorneZytkow}. Useful
references are also given in \cite{BisnoLamzin}. Thorne and
{\dotZ}ytkow wrote a very detailed paper \cite{ThorneZytkow} where
they built the first models of such objects.
%
The theory of Thorne--{\dotZ}ytkow objects was also studied in Refs.\
\cite{BisnoLamzin,Eich-ea,bbl01}.

Unfortunately, accurate modeling of Thorne--{\dotZ}ytkow objects is
extremely difficult. The main problem is to describe the energy
release, loss and transfer throughout the object, from its center, to the
outer layers of the neutron star, and through the rarefied convective
envelope. The most difficult place is near the bottom of the
convective envelope (close to the outer layers of the `inner' neutron
star). This layer is hot, neutrino emission carries away a lot of
energy, and convection is highly unusual. A small increase in the
neutrino losses may reduce the flow of heat from the neutron star to
the convective envelope, quench the convection, and the object will
be destroyed. Variations of thermal conductivity and nuclear reaction
rates under highly unusual conditions in these objects can also be
harmful. We add that there is no solid observational evidence that
these objects exist. Therefore, the existence of Thorne--{\dotZ}ytkow
objects remains an open question.

\section{Summary}
\label{Summary}

We have outlined the early history of the theory of neutron stars and
Landau's contribution to that theory (see also the chronology of
events in the Appendix). Our key point is that in Zurich, in February
1931, one year before the discovery of the neutron, Landau wrote a
remarkable paper \cite{Landau32}. In that paper he calculated the
maximum mass of white dwarf stars (independently of, but later than,
Chandrasekhar) and predicted the existence of dense stars which look
like giant atomic nuclei (a valid description of neutron stars
nowadays). He suggested also that the laws of quantum mechanics are
violated in these very dense stars. Let us be merciful with Landau
for that suggestion -- he was only 23 years old! Landau discussed his
paper with Bohr and Rosenfeld in Copenhagen in the period from 28
February to 19 March 1931, and published it in February 1932,  just
when the discovery of the neutron was announced. An accidental
coincidence of dates of the neutron's discovery and the publication
of Landau's paper produced confusion in the literature, with attempts
to claim that Landau predicted neutron stars just after learning
about the existence of the neutron. This claim seems definitely
invalid.

The question of whether Landau predicted neutron stars cannot be
answered unambiguously. On the one hand, yes, he predicted them, but
without neutrons and by violating quantum mechanics. On the other
hand, no, his vision of dense stars was too far from our knowledge
today. Perhaps, it would be more appropriate to say that he
superficially anticipated them. However, we do not insist and allow
the reader to judge. The next prediction of neutron stars by Baade
and Zwicky \cite{bz34a}  was certainly more realistic.

In 1937 Landau also contributed to the idea of stars with neutron
cores, which corresponded to what would later be called
Thorne--{\dotZ}ytkow objects. They are very difficult to model and
have not been observed.

After the car accident in 1962, Landau did not fully recover and,
being in constant pain, he could not concentrate on science. He died on
1 April 1968, just as the discovery of neutron stars as radio
pulsars, announced in {\it Nature} \cite{Hewish} on 24 February 1968,
was being widely discussed. Even if he was told about that discovery,
he was certainly not interested in it and could not understand the
importance of his contribution to the physics of compact stars. It
was our aim to remind the reader of this contribution and extract its
essence.

\paragraph{Acknowledgement.}
We greatly thank Finn Aaserud and Felicity Pors of the Niels Bohr
Archive in Copenhagen for tracing parts of this history.  We are also
grateful to Gennady Gorelik and Alexei Kojevnikov for useful
comments, and to Nicolas Chamel, Yurii N.\ Gnedin and Peter Shternin
for their help in finding old publications.

\section*{Appendix. Chronology}

\begin{enumerate}

\item
February 1931, Zurich. Landau finishes his paper \cite{Landau32}, in
which he calculates the maximum mass of white dwarfs and predicts the
existence of dense stars which look like giant atomic nuclei.

\item
25 February -- 19 March, 1931. Landau in Copenhagen. He most likely
discusses his paper with Bohr and Rosenfeld in the period from 28
February (when Rosenfeld arrives) to 19 March.

\item
7 January 1932. Landau submits his paper \cite{Landau32} to {\it
Physikalische Zeitschrift der Sowjetunion}.

\item
End of January 1932. Chadwick became interested in conducting the
experiment which led to the discovery of the neutron.

\item
17 February 1932. Chadwick submits his paper \cite{Chadwick} on the
discovery of the neutron to {\it Nature}.

\item
24 February 1932. Chadwick writes a letter to Bohr informing him of
the discovery of the neutron.

\item
27 February 1932. Chadwick's paper on the discovery of the neutron is
published.

\item
29 February 1932. Landau's paper \cite{Landau32} published.

\item
15--16 December 1933, Stanford. Baade and Zwicky give a talk at a
meeting of the American Physical Society suggesting the concept of
neutron stars, and their origin in supernova explosions.

\item
15 January 1934. The abstract of the talk by Baade and Zwicky is
published \cite{bz34a}.

\end{enumerate}


\end{document}